\documentclass[final,authoryear,3p,times,twocolumn]{elsarticle}

\usepackage{amssymb}

\journal{Applied Radiation and Isotopes}

\begin{document}

\begin{frontmatter}



\title{Investigation of activation cross-section data of proton induced nuclear reactions on rhenium}


\author[1]{F. Ditr\'oi\corref{*}}
\author[1]{F. T\'ark\'anyi}
\author[1]{S. Tak\'acs}
\author[2]{A. Hermanne}
\author[3]{H. Yamazaki} 
\author[3]{M. Baba}
\author[3]{A. Mohammadi} 
\author[4]{A.V.  Ignatyuk}
\cortext[*]{Corresponding author: ditroi@atomki.hu}

\address[1]{Institute of Nuclear Research of the Hungarian Academy of Sciences (ATOMKI),  Debrecen, Hungary}
\address[2]{Cyclotron Laboratory, Vrije Universiteit Brussel (VUB), Brussels, Belgium}
\address[3]{Cyclotron Radioisotope Center (CYRIC), Tohoku University, Sendai, Japan}
\address[4]{Institute of Physics and Power Engineering (IPPE), Obninsk, Russia}

\begin{abstract}
In the frame of systematic investigations of activation cross-section data for different applications the excitation functions of $^{nat}$Re(p,x)$^{185}$Os, $^{183m}$Os, $^{183g}$Os, $^{182}$Os, $^{181m}$Os, $^{186g}$Re, $^{184m}$Re, $^{184g}$Re, $^{183}$Re, $^{182m}$Re, $^{182g}$Re and $^{181g}$Re reactions were measured up to 70 MeV. The data for the $^{nat}$Re(p,x)$^{183m}$Os, $^{183g}$Os, $^{182}$Os, $^{181g}$Os, $^{186g}$Re, $^{184m}$Re,$^{182m}$Re, $^{182g}$Re, $^{181}$Re reactions are reported for the first time. Activation method, stacked foil irradiation technique and  $\gamma$-spectroscopy for activity detection were used. The experimental data were compared with predictions of three theoretical codes. From the measured cross-section thick target integral yields were also calculated and presented.
\end{abstract}


\end{frontmatter}


\section{Introduction}
\label{1}
To meet requirements for different practical applications we started to establish an experimental activation database some years ago by performing new experiments and a systematical survey of existing data of proton induced cross-sections up to 100 MeV and deuteron induced cross-sections up to 50 MeV \citep{TF2011,TF20112}.
The proton activation data for rhenium (Re) are relevant for accelerator and target technology (rhenium-tungsten alloy target in Fermilab, Austron, etc.); for medical radioisotope production; for producing radioactive ion beams (RIB); for controlled fusion experiments and reactors (ITER, DEMO, etc. ); for space applications; for thin layer activation (TLA); etc. 
Rhenium is a very heavy (atomic mass: 186.207 g/mol) and dense metal ($\rho$ = 21.02 g/cm$^3$), with high melting point (3180.0 $^o$C) and it is resistant to heat, wear and corrosion. Presently it is mostly used as an additive in super-alloys for aviation technology. The technical and medical applications are discussed in more detail in recently submitted work on activation data of deuteron induced reactions on rhenium (Ditrói et al., 2012).
Earlier literature contains only few experimental data on proton activation of rhenium. Armini and Bunker investigated the long-lived isotope production cross-sections from proton bombardment of rhenium, from 15 to 160 MeV, $^{185}$Os, $^{183}$Re(cum), $^{184g}$Re \citep{Armini}.
Dmitriev measured the thick target yield data for production of $^{185}$Os at 22 MeV in \citep{Dmitriev81} and \citep{Dmitriev83},  Ignatyuk \citep{Ignatyuk} and Okolovich \citep{Okolovich} investigated the properties of rhenium in the frame of a study on fissibility of sub-actinide nuclei by protons .
Results for estimation of production cross-sections on rhenium by protons through model calculations were produced by ALICE-IPPE code (Dityuk et al., 1998) in the MENDL-2p \citep{Shubin} database, with TALYS code \citep{Koning2007} in TENDL-2011 \citep{Koning2011} library and recently Maiti  \citep{Maiti}  published calculations for different light ion induced reactions for production of rhenium isotopes.

\begin{table*}[t]
\tiny
\caption{Decay characteristics of the investigated activation products and Q-values of contributing reactions}
\centering
\begin{center}
\begin{tabular}{|p{0.6in}|p{0.5in}|p{0.6in}|p{0.5in}|p{0.7in}|p{0.7in}|} \hline 
Nuclide & Half-life & E${}_{\gamma}$(keV) & I${}_{\gamma}$(\%) & Contributing reaction & Q-value\newline (keV) \\ \hline 
\textbf{${}^{185}$Os\newline }$\varepsilon $: 100 \% & 93.6 d & 646.116\newline 717.424 \newline 874.813\newline 880.523 & 78\newline 3.94 \newline 6.29\newline 5.17 & ${}^{185}$Re(p,n)\newline ${}^{187}$Re(p,3n) & -1795.144\newline -15331.34 \\ \hline 
\textbf{${}^{183m}$Os\newline }IT: 15\%\newline $\varepsilon $: 85\%\newline 170.71\textit{5 keV}  & 9.9 h & 1034.86\newline 1101.92\newline 1107.9 & 6.02\newline 49.0\newline 22.4 & ${}^{185}$Re(p,3n)\newline ${}^{187}$Re(p,5n) & -17084.4\newline ~-30620.6 \\ \hline 
\textbf{${}^{183g}$Os\newline }$\varepsilon $: 100 \% & 13.0 h & 114.47\newline 236.367\newline 381.763\newline 851.48 & 20.6\newline 3.41\newline 89.6\newline 4.56 & ${}^{185}$Re(p,3n)\newline ${}^{187}$Re(p,5n) & -17084.4\newline ~-30620.6 \\ \hline 
\textbf{${}^{182}$Os\newline }$\varepsilon $: 100\% & 21.84 h & 130.80\newline 180.20\newline 263.29 & 3.30\newline 34.1\newline 6.76 & ${}^{185}$Re(p,4n)\newline ${}^{187}$Re(p,6n) & -24209.4\newline ~-37745.6 \\ \hline 
\textbf{${}^{181g}$Os\newline } & 105~m & 238.75\newline 242.91~\newline 787.6~\newline 826.77~\newline 831.62~ & 44~\newline 6.1~\newline 5.3~\newline 20\newline 7.7~ & ${}^{185}$Re(p,5n)\newline ${}^{187}$Re(p,7n) & -33336.8\newline ~-46873.0 \\ \hline 
\textbf{${}^{186}$${}^{g}$Re\newline }$\beta $${}^{-}$: 92.53\%\newline $\varepsilon $: 7.47\% & 3.7183 d & 137.157 & 9.47 & ${}^{187}$Re(p,pn) & ~-7356.84 \\ \hline 
\textbf{${}^{184m}$Re\newline }IT: 74.5\%\newline $\varepsilon $: 25.5\%\newline 188.0463\textit{ keV} ~ & 169 d & 104.7395\newline 161.269\newline 216.547\newline 252.845\newline 318.008 & 13.6\newline 6.56\newline 9.5\newline 10.8\newline 5.81 & ${}^{185}$Re(p,pn)\newline ${}^{187}$Re(p,p3n)\newline  & -7666.84\newline -21203.03 \\ \hline 
\textbf{${}^{184g}$Re\newline }$\varepsilon $: 100\% & 35.4 d & 111.2174\newline 792.067\newline 894.760\newline 903.282 & 17.2\newline 37.7\newline 15.7\newline 38.1 & ${}^{185}$Re(p,pn)\newline ${}^{187}$Re(p,p3n) & -7666.84\newline -21203.03 \\ \hline 
\textbf{${}^{183}$Re} & 70.0 d\newline $\varepsilon $: 100 & 162.3266\newline 291.7282 & 23.3\newline 3.05 & ${}^{185}$Re(p,p2n)\newline ${}^{187}$Re(p,p4n)\newline ${}^{183}$Os decay & -14153.76\newline -27689.95 \\ \hline 
\textbf{${}^{182m}$Re\newline }$\varepsilon $: 100\%\newline 0+X \textit{keV} & 12.7 h & 100.12\newline 152.43 \newline 229. 32\newline 470.26*\newline 894.85*\newline 1121.4 \textit{\newline }1189.2\newline 1221.5\newline 1231.2 & 14.4\newline 7.0\newline 2.6\newline 2.02\newline 2.11\newline 32.0\newline 15.1\newline 25.0\newline 1.32 & ${}^{185}$Re(p,p3n)\newline ${}^{187}$Re(p,p5n) & -22589.0\newline ~-36125.0 \\ \hline 
\textbf{${}^{182g}$Re\newline }$\varepsilon $: 100\% & 64.0 h & 100.10\newline 130.81*\newline 169.15*\newline 191.39*\newline 229.32\newline 286.56*\newline 351.07*\newline 1076.2*\newline 1121.3\newline 1189.0\newline 1221.4\newline 1231.0\newline 1427.3*\underbar{} & 16.5\newline 7.5\newline 11.4\newline 6.7\newline 25.8\newline 7.1\newline 10.3\newline 10.6\newline  22.1\newline 9.1\newline 17.5\newline 14.9\newline 9.8 & ${}^{185}$Re(p,p3n)\newline ${}^{187}$Re(p,p5n)\newline  & -22589.0\newline ~-36125.0 \\ \hline 
\textbf{${}^{181}$Re\newline }$\varepsilon $: 100\%\textbf{} & 19.9 h & ~360.7\newline 365.5 & 20\newline ~56 & ${}^{185}$Re(p,p4n)\newline ${}^{187}$Re(p,p6n) & -29596.0\newline -43132.2 \\ \hline 
\end{tabular}

\end{center}
\begin{flushleft}
\footnotesize{\noindent The Q-values refer to formation of the ground state and are obtained from \citep{Pritychenko}.

\noindent When complex particles are emitted instead of individual protons and neutrons the Q-values have to be decreased by the respective binding energies of the compound particles: np-d, +2.2 MeV; 2np-t, +8.48 MeV; n2p-${}^{3}$He, +7.72 MeV; 2n2p-$\alpha$, +28.30 MeV
\noindent The independent $\gamma$-lines are marked with ``*''}
\end{flushleft}

\end{table*}

\section{Experiment and data evaluation}
\label{2}
The excitation functions were measured up to 70 MeV incident energy via the activation technique by bombarding stacked Mo(52.5\%)Re(47.5\%) alloy foils ( \textcopyright Goodfellow \textgreater 99.98\%, thickness 50 $\mu$m) targets with a low intensity proton  beam at the AVF-930 cyclotron of the Cyclotron Laboratory (CYRIC) of the Tohoku University, Sendai, Japan and with 37 MeV protons at the CGR-560 cyclotron of the Vrije Universiteit Brussel (VUB), Brussels, Belgium. 
The procedures for irradiation, activity measurement and the data evaluation method (including estimation of uncertainties) were similar as described in several earlier works published by authors from the research groups in Debrecen, Brussels and Sendai. Here we summarize only the most salient features of the experimental and data evaluation technique pertaining to the present work \citep{Ditroi}. 
Special care was taken in preparation of uniform targets with known thickness, in determination of the energy and of the intensity of the bombarding beam along the target stack and in determination of the activities of the samples. The target foils were purchased from the Goodfellow company with guaranteed thickness tolerances. The sizes and weights of the whole metal sheets were measured and the average thickness was calculated. After cutting them into proper sizes they were measured again and the uniformity was verified by comparing them to the average values.
In case of the CYRYC irradiation the MoRe target foils were stacked together with other target foils (CuMnNi alloy, ScO, NdO) and with 100 $\mu$m thick Al foils used to monitor the beam parameters. The irradiation of the targets was performed in He gas atmosphere in a water cooled target holder. The target stack was irradiated with a collimated, 70 MeV incident energy proton beam for 20 min at about 25 nA. 
The radioactivity of each sample and monitor foil was measured non-destructively by HPGe $\gamma$-spectrometry. Counting was started about 40 hours after the end of the bombardment (EOB). 
For the VUB irradiation the stack contained 50 $\mu$m MoRe target foils and 12 $\mu$m Ti monitor foils. The stack was irradiated in a Faraday-cup like target holder, equipped with a collimator (effective beam diameter on target is 5 $\mu$m) and a secondary electron suppressor. Irradiation took place with a collimated, 36 MeV incident energy proton beam for 30 min at a constant beam current of 130 nA. Counting of the samples started about 6 hours after EOB. 
The decay data were taken from NUDAT \citep{Kinsey} (see Table 1), the reaction Q-values from Pritychenko and A. Sonzogni \citep{Pritychenko} the standard cross-section data for the used monitor reactions $^{nat}$Al(p,x)$^{22,24}$Na and $^{nat}$Ti(p,x)$^{48}$V were taken from T\'ark\'anyi \citep{TF2001} updated in 2006 \citep{IAEA}.
The energy degradation along the stack was determined via calculation \citep{Andersen} and corrected on the basis of the simultaneously measured monitor reactions by the method described by Tárkányi \citep{TF1991}. The excitation functions were hence determined in an accurate way relative to the simultaneously re-measured monitor reactions. (see Fig. 1).

\begin{figure}
\includegraphics[width=0.5\textwidth]{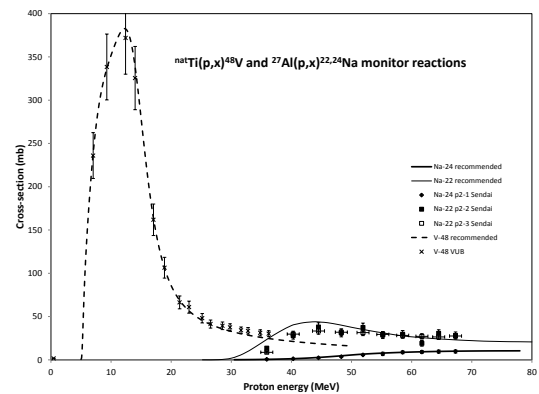}
\caption{Application of $^{nat}$T(p,x)$^{48}$V and $^{27}$Al(p,x)$^{22,24}$Na monitor reactions for determination of proton  beam energy and intensity}
\label{fig:1}       
\end{figure}

As naturally occurring rhenium is composed of two stable isotopes $^{185}$Re (37.40\%), and  $^{187}$Re (62.60\%), so called elemental cross-sections were determined supposing the rhenium is composed of only one single isotope. 
The resulting uncertainty on the cross-sections contain the individual uncertainties of the processes contributing linearly to the final result and are calculated accordingly to the well-accepted summation rules \citep{Hiba}. Quadratic summation of the uncertainties of the contributing parameters results an absolute error of 8.5\% (target thickness (2\%), detector efficiency (5\%), physical decay data (5\%), integrated beam current (2\%), counting statistics (2\%)). The uncertainties of the non-linear processes like half-life, irradiation time and measuring time were neglected. 
The uncertainty on the median energy in the targets was estimated from the uncertainty of the energy of the bombarding beam (starting from 0.3 MeV by the incident energy), the uncertainty in the thickness and uniformity of all foils, the beam straggling taking into account the cumulative effects.

\section{Tehoretical calculations}
\label{3}
The measured excitation functions were compared to theoretical effective cross-sections calculated by means of three different computer codes. For the pre-compound model codes ALICE-IPPE \citep{Dityuk} and EMPIRE-II \citep{Herman} the parameters for the optical model, level densities and pre-equilibrium contributions were taken as described in \citep{Belgya}. The third set of values in figures represents data in the TENDL 2011 online library \citep{Koning2011} calculated with the 1.4 version of TALYS. Results of ALICE-IPPE for excited states were obtained by applying the isomeric ratios derived from the EMPIRE code to the total cross-sections from ALICE.
For each activation product reaction cross-sections on the individual target isotopes were calculated up to 100 MeV proton energy and a weighted summation (weighting factor is abundance of natural occurrence) was made to obtain the production cross-section. In case of $^{185}$Os, $^{183}$Re and $^{184g}$Re the TENDL 2011 data \citep{Koning2011} are presented up to 200 MeV for comparison with the earlier experimental data of Armini \citep{Armini}. 

\section{Results and discussions}
\label{4}

\subsection{Cross-sections}
\label{4.1}
By irradiating rhenium with 70 MeV protons, many radioisotopes of different elements are produced in significant amounts. Among the radionuclides formed, unfortunately many have decay characteristics (very short or too long half-life or no $\gamma$-lines) resulting in activities below the detection limits in our measuring circumstances (long waiting time in the high energy irradiation). In the 2 experiments only radioisotopes of Os and Re were identified with proper reliability. 
The measured experimental cross-section data are shown in Figs 2-17 together with the theoretical results. The contributions of the two stable target isotopes ($^{185}$Re and $^{187}$Re) are well separated. The numerical values essential for further evaluation are collected in Tables 2-4. In the figures the results of irradiations made at different primary energies at different accelerators are marked separately. The results of the two irradiations show good agreement in the overlapping energy range.

\subsubsection{Production of osmium isotopes}
\label{4.1.1}
The radioisotopes of osmium are produced only directly by (p,xn) reactions. In our spectra we could identify the $^{185}$Os, $^{183m}$Os, $^{183g}$Os, $^{182}$Os radioisotopes. Due to the long waiting time after EOB we could not find the relatively longer-lived $^{181}$Os (105 min) in the measurements after the 70 MeV irradiation.

\vspace{ 2 mm}
\textbf{$^{nat}$Re(p,x)$^{185}$Os}

The $^{185}$Os radioisotope (93.6 d) is formed by the $^{185}$Re(p,n)$^{185}$Os  and $^{187}$Re(p,3n)$^{185}$Os  reactions.  The contributions of the two reactions are well separated in Fig. 2. The maximum of the (p,n) cross-section is significantly lower than that for (p,3n). Our data are in acceptable agreement with the earlier experimental data of Armini \citep{Armini}. The TALYS and ALICE codes reproduce well the low energy part, but underestimate the high energy tail. EMPIRE predicts a significantly wider $^{187}$Re(p,3n) peak than we found experimentally.

\begin{figure}
\includegraphics[width=0.5\textwidth]{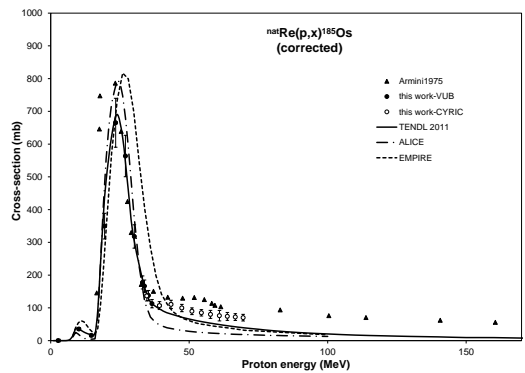}
\caption{Experimental and theoretical excitation functions for $^{nat}$Re(p,x)$^{185}$Os reaction}
\label{fig:2}       
\end{figure}

\vspace{ 2 mm}
\textbf{$^{nat}$Re(p,x)$^{183m}$Os and $^{nat}$Re(p,x)$^{183g}$Os}
A shorter-lived excited state (T$_{1/2}$ = 9.9 h) and a longer-lived (13.0 h) ground state exist for $^{183}$Os. The isomeric state has a 15\% isomeric transition probability to the ground state. We present here independent results for production of $^{183m}$Os (Fig. 3) and for $^{183g}$Os, corrected for the isomeric decay of $^{183m}$Os (Fig. 4). The shape of the theoretical predictions follows our experimental results, but the cross-section values for $^{183m}$Os production are underestimated by the TENDL 2011, while for $^{183g}$Os there are overestimations in all codes for the low energy maximum.  At the high energy maximum the agreement is better. The width of the peaks of the calculated $^{185}$Re(p,3n), $^{187}$Re(p,5n) contributions in the case of EMPIRE is larger in both cases. 

\begin{figure}
\includegraphics[width=0.5\textwidth]{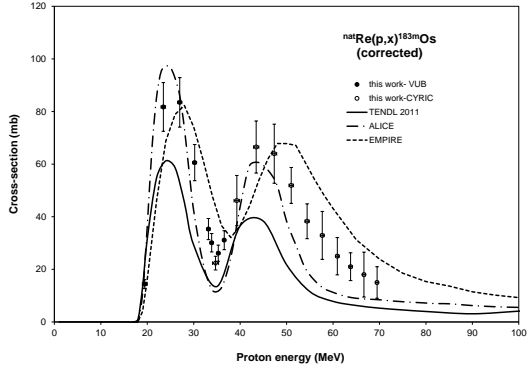}
\caption{Experimental and theoretical excitation functions for $^{nat}$Re(p,x)$^{183m}$Os reaction}
\label{fig:3}       
\end{figure}

\begin{figure}
\includegraphics[width=0.5\textwidth]{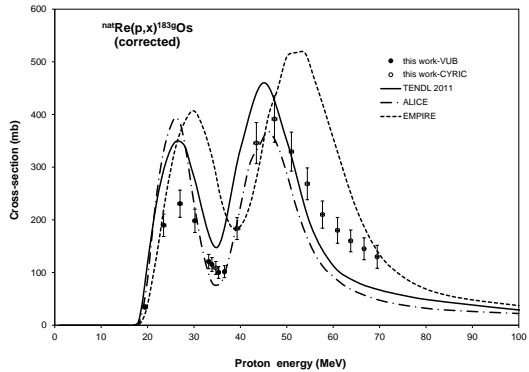}
\caption{Experimental and theoretical excitation functions for $^{nat}$Re(p,x)$^{183g}$Os reaction}
\label{fig:4}       
\end{figure}

\vspace{ 2 mm}
\textbf{$^{nat}$Re(p,x)$^{182}$Os}
The experimental and theoretical data for production of the $^{182}$Os (21.84 h) are shown in Fig. 5. There is a good agreement between the experimental data and the TENDL 2011 results in the whole investigated energy range. ALICE-IPPE overestimates peak values by 50\% for (p,4n) but agrees well for the (p,6n). The broader peaks and an energy shift of about 10 MeV for EMPIRE predictions can be observed.

\begin{figure}
\includegraphics[width=0.5\textwidth]{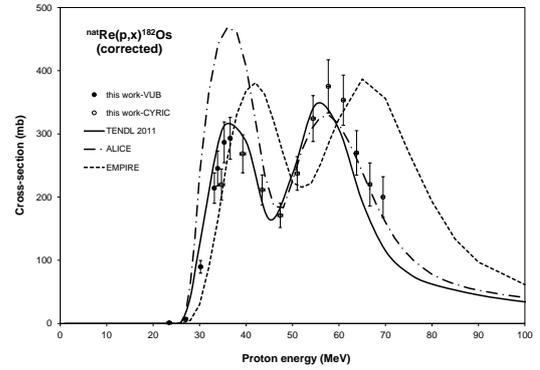}
\caption{Experimental and theoretical excitation functions for $^{nat}$Re(p,x)$^{182}$Os reaction}
\label{fig:5}       
\end{figure}

\vspace{ 2 mm}
\textbf{$^{nat}$Re(p,x)$^{181g}$Os}
Our two experimental data points and the theoretical data for production of the $^{181g}$Os (105 min) are shown in Fig. 6. It has no isomeric decay from the short-lived excited state (2.7 min). In spite of the predictions of the theoretical calculations we could not detect $^{181g}$Os in the high energy irradiation, due to its short half-life and the long cooling time. The EMPIRE results are energy shifted, compared to the predictions of the two other codes.

\begin{figure}
\includegraphics[width=0.5\textwidth]{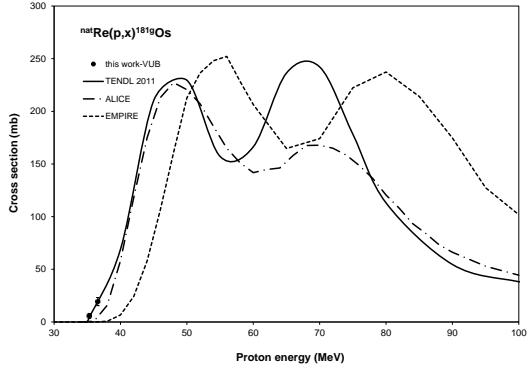}
\caption{Experimental and theoretical excitation functions for $^{nat}$Re(p,x)$^{181}$Os reaction}
\label{fig:6}       
\end{figure}

\subsubsection{Cross-sections of residual radio-products of rhenium}
\label{4.1.2}
The radioisotopes of Re can be formed in two routes: directly via (p,pxn) reactions and by decay of simultaneously produced progenitor radioisotopes of Os and W. The effect of the secondary neutrons via (n,$\gamma$) and (n,2n) reactions should also be taken into account, but it  was negligibly small in our experiments according to measurement of produced radioisotopes behind the range of the bombarding charged particles.

\vspace{ 2 mm}
\textbf{$^{nat}$Re(p,x)$^{186g}$Re}
The $^{186g}$Re (3.7183 d) is produced directly only through the $^{187}$Re(p,pn) reaction and through the internal decay (IT/0.150) of the very long half-life 186mRe(2.0 10$^5$ a), which contribution is negligible in the present experiment. The measured experimental results for $^{186g}$Re are shown in Fig. 7 in comparison with the theoretical data. The model codes predict acceptable well the experimental results (both in shape and in magnitude).

\begin{figure}
\includegraphics[width=0.5\textwidth]{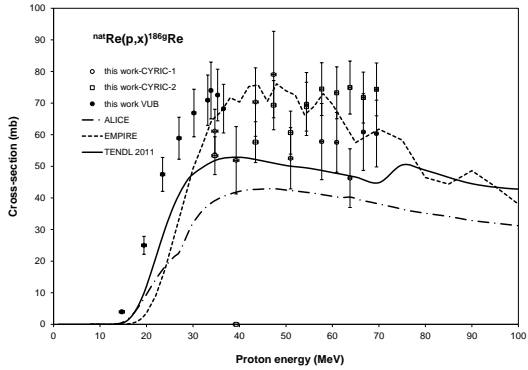}
\caption{Experimental and theoretical excitation functions for $^{nat}$Re(p,x)$^{186g}$Re reaction}
\label{fig:7}       
\end{figure}

\vspace{ 2 mm}
\textbf{$^{nat}$Re(p,x)$^{184m}$Re and $^{nat}$Re(p,x)$^{184g}$Re}
The $^{184}$Re radionuclide has a long-lived (T$_{1/2}$ =169 d) metastable state decaying for 74.5\% to the 35.4 d ground state by IT. According to Fig. 8 the measured excitation function for $^{184m}$Re (no clear distinction in contribution of the 2 stable target isotopes) is in good agreement with the results of ALICE and EMPIRE calculations, but it significant lower than the data in the TENDL 2011 library.
We present here independent results for production of $^{184g}$Re (35.4 d), which does not contain the contribution from the decay of isomeric state. For comparison also the values published by Armani \citep{Armini} are presented. The agreement of the experimental data and the theory in this case is much better (see Fig. 9) also for TENDL 2011 data.

\begin{figure}
\includegraphics[width=0.5\textwidth]{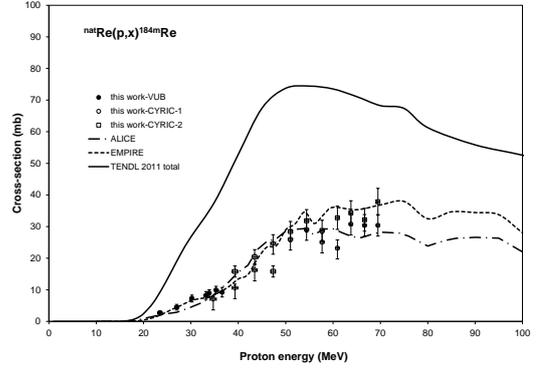}
\caption{Experimental and theoretical excitation functions for $^{nat}$Re(p,x)$^{184m}$Re reaction}
\label{fig:8}       
\end{figure}

\begin{figure}
\includegraphics[width=0.5\textwidth]{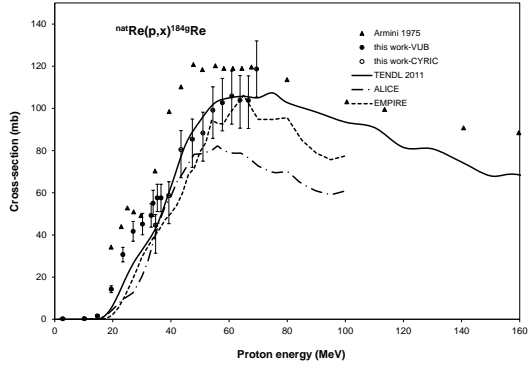}
\caption{Experimental and theoretical excitation functions for $^{nat}$Re(p,x)$^{184g}$Re reaction}
\label{fig:9}       
\end{figure}

\vspace{ 2 mm}
\textbf{$^{nat}$Re(p,x)$^{183}$Re}
The measured cumulative cross-sections of the $^{183}$Re (70.0 d) contain contribution from the $^{185}$Re(p,p2n) and $^{187}$Re(p,p4n) reactions and from the $^{183m}$Os (9.9 h) and $^{183g}$Os (13.0 h) decay. The data were obtained from  -spectra measured after a long cooling time. As it shown in Fig. 10 the theories predict the magnitude and the shape of the excitation function more or less properly. 

\begin{figure}
\includegraphics[width=0.5\textwidth]{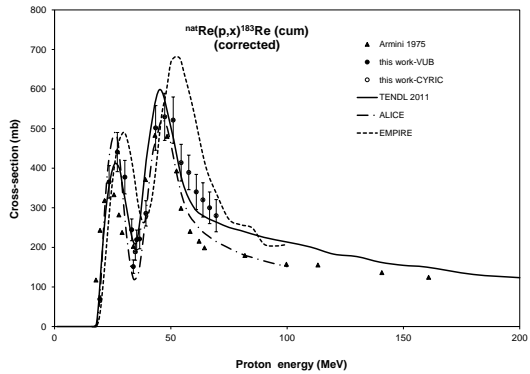}
\caption{Experimental and theoretical excitation functions for $^{nat}$Re(p,x)$^{183}$Re (cum) reaction}
\label{fig:10}       
\end{figure}

\vspace{ 2 mm}
\textbf{$^{nat}$Re(p,x)$^{182m}$Re and $^{nat}$Re(p,x)$^{182g}$Re}
The radionuclide 182Re has two longer-lived isomeric states, both decay independently. The  higher energy, low spin isomeric state (12.7 h, 2$^+$) is produced directly and through the decay of the parent $^{182}$Os (21.84 h), while  $^{182g}$Re (64 h, 7$^+$) is only produced directly. In Fig. 11 we present independent cross-sections for the metatable state after substracting the conribution of the parent decay. The correction was possible only for the low energy irradiation and the correction is so large that the final data contains very large uncertainties. For the high energy irradiation  the first measurerement started  only after 3 half-life of $^{182m}$Re, resulting in very low count rates of this radioisotope. The new experimental data for production of the $^{182g}$Re are shown in Fig. 12.

\begin{figure}
\includegraphics[width=0.5\textwidth]{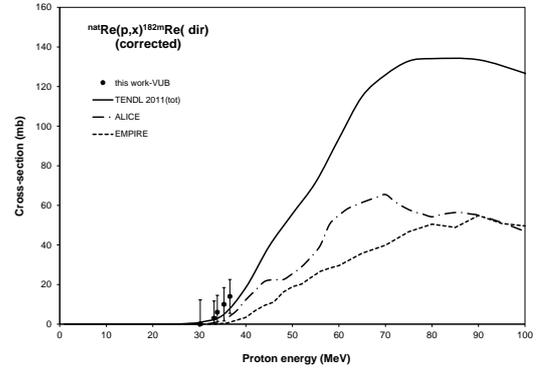}
\caption{Experimental and theoretical excitation functions for $^{nat}$Re(p,x)$^{182m}$Re reaction}
\label{fig:11}       
\end{figure}

\begin{figure}
\includegraphics[width=0.5\textwidth]{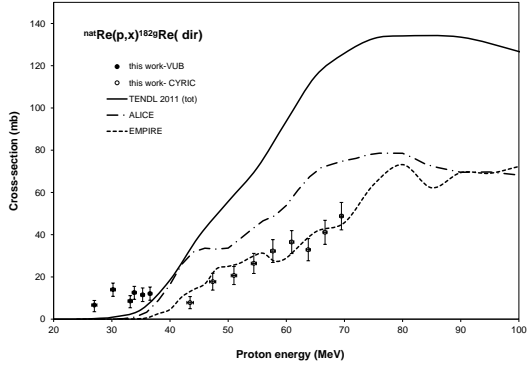}
\caption{Experimental and theoretical excitation functions for $^{nat}$Re(p,x)$^{182g}$Re reaction}
\label{fig:12}       
\end{figure}

\vspace{ 2 mm}
\textbf{$^{nat}$Re(p,x)$^{181}$Re}
The measured cumulative cross-sections contain the full contribution from decay of both isomeric states of the parent $^{181}$Os (half-lives are 105 min and 2.7 min respectively). The agreement with the results of the theoretical codes is acceptably good. 

\begin{figure}
\includegraphics[width=0.5\textwidth]{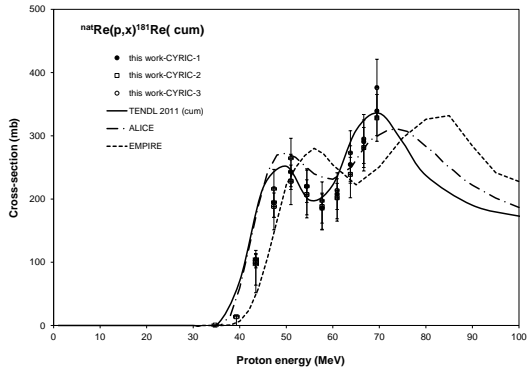}
\caption{Experimental and theoretical excitation functions for $^{nat}$Re(p,x)$^{181m}$Re reaction}
\label{fig:13}       
\end{figure}

\begin{table*}[t]
\tiny
\caption{Measured cross-sections for production of the ${}^{185}$Os,${}^{183m}$Os, ${}^{183g}$Os, ${}^{182}$Os${}^{ }$and ${}^{181g}$Os radionuclides}
\centering
\begin{center}
\begin{tabular}{|p{0.4in}|p{0.4in}|p{0.4in}|p{0.4in}|p{0.4in}|p{0.4in}|p{0.4in}|p{0.4in}|p{0.4in}|p{0.4in}|p{0.4in}|p{0.4in}|} \hline 
\multicolumn{2}{|p{0.8in}|}{\textbf{Energy ${\mathbf ±}$ ${\mathbf \Delta }$E\newline (MeV) }} & \multicolumn{10}{|p{4.in}|}{\textbf{Cross-section ${\mathbf ±}$ ${\mathbf \Delta }$${\mathbf \delta }$ (mbarn)}} \\ \hline 
\multicolumn{2}{|p{0.8in}|}{\textbf{}} & \multicolumn{2}{|p{0.8in}|}{${}^{185}$Os} & \multicolumn{2}{|p{0.9in}|}{${}^{183m}$Os} & \multicolumn{2}{|p{0.8in}|}{${}^{183g}$Os} & \multicolumn{2}{|p{0.8in}|}{${}^{182}$Os} & \multicolumn{2}{|p{0.8in}|}{${}^{181g}$Os} \\ \hline 
\multicolumn{12}{|p{1in}|}{\textbf{VUB}} \\ \hline 
36.6 & 0.3 & 112.8 & 12.7 & 31.1 & 3.6 & 101.6 & 11.4 & 293.1 & 33.0 & 19.3 & 3.8 \\ \hline 
35.3 & 0.3 & 137.6 & 15.5 & 26.1 & 3.1 & 100.1 & 11.3 & 286.7 & 32.2 & 5.7 & 2.4 \\ \hline 
33.9 & 0.3 & 166.3 & 18.7 & 30.1 & 3.5 & 115.2 & 13.0 & 245.4 & 27.6 &  &  \\ \hline 
33.2 & 0.4 & 177.7 & 20.0 & 35.3 & 4.0 & 120.8 & 13.6 & 214.1 & 24.1 &  &  \\ \hline 
30.2 & 0.4 & 317.9 & 35.7 & 60.6 & 6.9 & 198.2 & 22.3 & 89.6 & 10.1 &  &  \\ \hline 
27.0 & 0.4 & 563.6 & 63.3 & 83.5 & 9.4 & 230.7 & 25.9 & 6.9 & 1.1 &  &  \\ \hline 
23.5 & 0.5 & 665.2 & 74.7 & 81.7 & 9.3 & 189.6 & 21.3 & 1.0 & 1.3 &  &  \\ \hline 
19.5 & 0.5 & 349.6 & 39.2 & 14.4 & 1.7 & 34.5 & 3.9 &  &  &  &  \\ \hline 
14.7 & 0.6 & 15.7 & 1.8 &  &  &  &  &  &  &  &  \\ \hline 
10.2 & 0.6 & 35.4 & 4.0 &  &  &  &  &  &  &  &  \\ \hline 
2.8 & 0.6 & 0.1 & 0.0 &  &  &  &  &  &  &  &  \\ \hline 
\multicolumn{10}{|p{1in}|}{\textbf{CYRIC}} & \textbf{} & \textbf{} \\ \hline 
69.5 & 0.3 & 70.0 & 10.3 & 15.0 & 6.0 & 130.0 & 21.9 & 200.0 & 32.3 &  &  \\ \hline 
66.7 & 0.3 & 72.0 & 10.3 & 18.0 & 8.5 & 145.0 & 20.8 & 220.0 & 34.2 &  &  \\ \hline 
63.8 & 0.3 & 74.0 & 12.1 & 21.0 & 5.3 & 160.0 & 20.7 & 270.0 & 35.2 &  &  \\ \hline 
60.9 & 0.4 & 76.0 & 15.9 & 25.0 & 7.1 & 180.0 & 24.5 & 353.4 & 39.7 &  &  \\ \hline 
57.7 & 0.4 & 80.0 & 12.2 & 32.9 & 9.1 & 210.0 & 25.8 & 375.2 & 42.1 &  &  \\ \hline 
54.4 & 0.4 & 84.8 & 9.7 & 38.3 & 6.6 & 268.3 & 30.2 & 324.3 & 36.4 &  &  \\ \hline 
51.0 & 0.5 & 89.7 & 10.3 & 51.9 & 6.8 & 329.5 & 37.0 & 237.3 & 26.6 &  &  \\ \hline 
47.4 & 0.5 & 99.5 & 11.3 & 64.0 & 11.2 & 391.3 & 44.1 & 171.0 & 19.2 &  &  \\ \hline 
43.5 & 0.6 & 110.6 & 12.5 & 66.5 & 9.9 & 345.8 & 38.9 & 211.4 & 23.7 &  &  \\ \hline 
39.3 & 0.6 & 106.9 & 12.1 & 46.1 & 9.6 & 183.5 & 20.8 & 268.5 & 30.1 &  &  \\ \hline 
34.7 & 0.6 & 136.5 & 15.5 & 22.4 & 2.6 & 108.7 & 12.2 & 219.8 & 24.7 &  &  \\ \hline 
\end{tabular}

\end{center}
\end{table*}

\begin{table*}[t]
\tiny
\caption{Measured cross-sections for production of the ${}^{186}$Re, ${}^{184m}$Re, ${}^{184g}$Re, ${}^{183}$Re, ${}^{182m}$Re, ${}^{182g}$Re and${}^{ 181}$Re radionuclides}
\centering
\begin{center}
\begin{tabular}{|p{0.2in}|p{0.2in}|p{0.4in}|p{0.2in}|p{0.4in}|p{0.2in}|p{0.4in}|p{0.2in}|p{0.4in}|p{0.2in}|p{0.4in}|p{0.2in}|p{0.4in}|p{0.2in}|p{0.4in}|p{0.2in}|} \hline 
\multicolumn{2}{|p{0.4in}|}{\textbf{Energy ${\mathbf ±}$ ${\mathbf \Delta }$E}} & \multicolumn{14}{|p{4.2in}|}{\textbf{Cross-sections ${\mathbf ±}$ ${\mathbf \Delta }$${\mathbf \delta }$ (mbarn)}} \\ \hline 
\multicolumn{2}{|p{0.4in}|}{\textbf{(MeV)}} & \multicolumn{2}{|p{0.6in}|}{${}^{1}$${}^{86}$${}^{g}$Re} & \multicolumn{2}{|p{0.6in}|}{ ${}^{184m}$Re} & \multicolumn{2}{|p{0.6in}|}{${}^{184g}$Re} & \multicolumn{2}{|p{0.6in}|}{${}^{183}$Re} & \multicolumn{2}{|p{0.6in}|}{${}^{182m}$Re} & \multicolumn{2}{|p{0.6in}|}{${}^{182g}$Re} & \multicolumn{2}{|p{0.6in}|}{${}^{181}$Re} \\ \hline 
\multicolumn{16}{|p{0.6in}|}{\textbf{VUB}} \\ \hline 
36.6 & 0.30 & 68.2 & 7.7 & 9.2 & 1.5 & 57.5 & 6.5 & 221.2 & 25.0 & 14.0 & 9.1 & 12.1 & 3.2 &  &  \\ \hline 
35.3 & 0.32 & 72.5 & 8.2 & 9.9 & 1.2 & 57.6 & 6.5 & 219.2 & 24.6 & 10.0 & 10.6 & 11.5 & 3.3 &  &  \\ \hline 
33.9 & 0.35 & 74.0 & 9.0 & 9.0 & 1.1 & 55.0 & 6.2 & 151.3 & 17.0 & 6.0 & 9.2 & 12.6 & 3.0 &  &  \\ \hline 
33.2 & 0.38 & 70.9 & 8.0 & 8.2 & 1.2 & 49.2 & 5.6 & 244.3 & 27.5 & 3.0 & 6.1 & 8.5 & 2.6 &  &  \\ \hline 
30.2 & 0.41 & 66.9 & 7.5 & 7.3 & 1.1 & 45.2 & 5.1 & 377.3 & 42.4 &  &  & 14.0 & 3.1 &  &  \\ \hline 
27.0 & 0.44 & 58.9 & 6.6 & 4.5 & 0.9 & 41.7 & 4.7 & 440.7 & 49.5 &  &  & 6.7 & 2.2 &  &  \\ \hline 
23.5 & 0.48 & 47.5 & 5.4 & 2.7 & 0.7 & 30.7 & 3.5 & 365.3 & 41.1 &  &  & ~ &  &  &  \\ \hline 
19.5 & 0.51 & 25.0 & 2.8 & ~ &  & 14.3 & 1.6 & 69.0 & 7.8 &  &  & ~ &  &  &  \\ \hline 
14.7 & 0.56 & 4.0 & 0.5 & ~ &  & 1.6 & 0.2 & ~ &  &  &  & ~ &  &  &  \\ \hline 
10.2 & 0.60 &  &  & ~ &  & 0.3 & 0.1 & ~ &  &  &  & ~ &  &  &  \\ \hline 
2.8 & 0.65 &  &  & ~ &  & 0.3 & 0.1 & ~ &  &  &  & ~ &  &  &  \\ \hline 
\multicolumn{16}{|p{0.6in}|}{\textbf{CYRIC-1}} \\ \hline 
69.5 & 0.30 & 60.4 & 10.5 & 30.4 & 3.4 & 118.7 & 13.3 & 280.0 & 40.7 &  &  & 48.8 & 6.5 & 339.3 & 39.0 \\ \hline 
66.7 & 0.32 & 60.9 & 12.1 & 30.4 & 3.4 & 103.8 & 11.7 & 300.0 & 40.0 &  &  & 41.1 & 5.7 & 290.7 & 34.4 \\ \hline 
63.8 & 0.35 & 46.3 & 9.3 & 30.8 & 3.5 & 103.9 & 11.7 & 320.0 & 43.3 &  &  & 32.9 & 5.3 & 254.6 & 29.6 \\ \hline 
60.9 & 0.38 & 57.6 & 9.5 & 23.2 & 2.6 & 105.9 & 11.9 & 340.0 & 44.5 &  &  & 36.5 & 5.4 & 207.7 & 24.3 \\ \hline 
57.7 & 0.41 & 57.8 & 12.1 & 25.1 & 2.8 & 102.7 & 11.5 & 389.3 & 43.7 &  &  & 32.2 & 5.4 & 185.0 & 23.0 \\ \hline 
54.4 & 0.44 & 69.7 & 10.0 & 29.1 & 3.3 & 99.1 & 11.1 & 413.9 & 46.5 &  &  & 26.4 & 4.7 & 220.2 & 25.6 \\ \hline 
51.0 & 0.48 & 52.5 & 9.7 & 26.0 & 2.9 & 88.3 & 9.9 & 521.6 & 58.5 &  &  & 20.7 & 4.3 & 242.9 & 27.9 \\ \hline 
47.4 & 0.51 & 79.0 & 13.7 & 24.7 & 2.8 & 85.4 & 9.6 & 529.9 & 59.5 &  &  & 17.7 & 4.0 & 194.9 & 24.0 \\ \hline 
43.5 & 0.56 & 70.3 & 10.8 & 16.2 & 1.8 & 80.5 & 9.0 & 501.9 & 56.3 &  &  & 7.8 & 2.9 & 104.9 & 13.8 \\ \hline 
39.3 & 0.60 & 51.9 & 10.7 & 10.6 & 1.2 & 58.7 & 6.6 & 285.8 & 32.1 &  &  &  &  & ~ &  \\ \hline 
34.7 & 0.65 & 61.1 & 7.0 & 7.1 & 0.8 & 44.7 & 5.0 & 188.8 & 21.2 &  &  &  &  & ~ &  \\ \hline 
\multicolumn{16}{|p{0.6in}|}{\textbf{CYRIC-2}} \\ \hline 
69.5 & 0.30 & 74.3 & 8.3 & 37.9 & 4.2 &  &  &  &  &  &  &  &  & 328.3 & 37.0 \\ \hline 
66.7 & 0.32 & 71.7 & 8.1 & 32.1 & 3.6 &  &  &  &  &  &  &  &  & 281.4 & 31.8 \\ \hline 
63.8 & 0.35 & 74.9 & 8.4 & 34.3 & 3.9 &  &  &  &  &  &  &  &  & 239.0 & 26.9 \\ \hline 
60.9 & 0.38 & 73.3 & 8.2 & 32.8 & 3.7 &  &  &  &  &  &  &  &  & 201.9 & 22.8 \\ \hline 
57.7 & 0.41 & 74.4 & 8.4 & 28.8 & 3.2 &  &  &  &  &  &  &  &  & 188.2 & 21.3 \\ \hline 
54.4 & 0.44 & 68.9 & 7.7 & 31.8 & 3.6 &  &  &  &  &  &  &  &  & 207.1 & 23.3 \\ \hline 
51.0 & 0.48 & 60.7 & 6.8 & 28.4 & 3.2 &  &  &  &  &  &  &  &  & 228.1 & 25.7 \\ \hline 
47.4 & 0.51 & 69.3 & 7.8 & 15.8 & 1.8 &  &  &  &  &  &  &  &  & 188.2 & 21.3 \\ \hline 
43.5 & 0.56 & 57.7 & 6.5 & 20.4 & 2.3 &  &  &  &  &  &  &  &  & 100.7 & 11.4 \\ \hline 
39.3 & 0.60 &  &  & 15.8 & 1.8 &  &  &  &  &  &  &  &  & 13.8 & 2.2 \\ \hline 
34.7 & 0.65 & 53.4 & 6.0 & ~ &  &  &  &  &  &  &  &  &  & 0.5 & 0.2 \\ \hline 
\end{tabular}

\end{center}
\end{table*}

\subsection{Integral yields}
\label{4.2}
On the basis of the measured experimental data we calculated the integral yields for the proton induced reactions on $^{nat}$Re (see Fig. 14 and Fig. 15). Only one experimental work reporting on thick target yield data was found in the literature for production of $^{185}$Os at 22 MeV \citep{Dmitriev81}, which corresponds well to our new results. 

\begin{figure}
\includegraphics[width=0.5\textwidth]{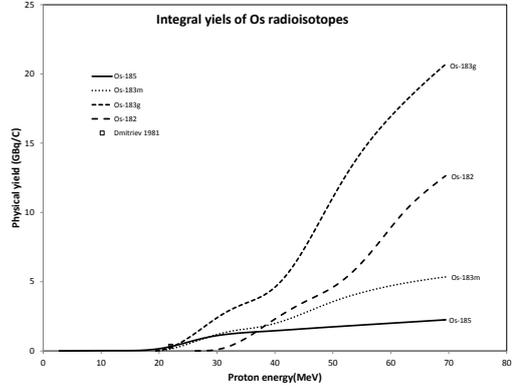}
\caption{Integral yields for production of some Os radioisotopes}
\label{fig:14}       
\end{figure}

\begin{figure}
\includegraphics[width=0.5\textwidth]{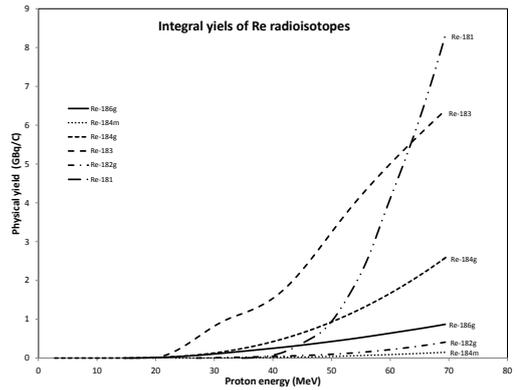}
\caption{Integral yields for production of some Re radioisotopes}
\label{fig:15}       
\end{figure}

\section{Summary and conclusion}
\label{5}
Activation cross-sections of proton induced nuclear reactions on rhenium were measured for 12 reaction products up to 70 MeV, out of them data for 7 reactions are presented here for the first time. Model calculations were done by using the EMPIRE and Alice-IPPE codes. The results were also compared with the data of the TALYS based TENDL 2011 online library. The predictions of theoretical calculations could be considered only moderately successful, especially for isomeric states. The obtained experimental data provide a basis for model calculations and for different applications.

\section{Acknowledgements}
\label{6}

This work was done in the frame MTA-FWO research project and ATOMKI-CYRIC collaboration. The authors acknowledge the support of research projects and of their respective institutions in providing the materials and the facilities for this work. 
 



\clearpage
\bibliographystyle{elsarticle-harv}
\bibliography{Rep}







\end{document}